\documentclass[nofootinbib,prd,11pt,superscriptaddress]{revtex4}%
\usepackage{amsmath}
\usepackage{amsfonts}
\usepackage{amssymb}
\usepackage{graphicx}%
\usepackage{amsmath,amssymb}
\usepackage{mathrsfs}
\usepackage{graphicx}
\usepackage{color}
\usepackage{subfigure}
\usepackage{fancyhdr}
\usepackage{multirow}
\usepackage{float}
\usepackage{epsfig}
\usepackage{amsfonts}
\usepackage{bm}

\def\be{\begin{equation}}
\def\ee{\end{equation}}
\def\beq{\begin{eqnarray}}
\def\eeq{\end{eqnarray}}

\begin{document}

\title{Traversable Wormholes in Minimally Geometrical Deformed \\Trace-Free Gravity using Gravitational Decoupling}

\author{Piyachat Panyasiripan}
\email{piyachat.pa@mail.wu.ac.th}
\affiliation{School of Science, Walailak University, Nakhon Si Thammarat, 80160, Thailand}

\author{Narakorn Kaewkhao}
\email{naragorn.k@psu.ac.th}
\affiliation{Department of Physics, Faculty of Science, Prince of Songkla University, Hatyai 90112, Thailand} 

\author{Phongpichit Channuie}
\email{phongpichit.ch@mail.wu.ac.th}
\affiliation{School of Science, Walailak University, Nakhon Si Thammarat, 80160, Thailand}
\affiliation{College of Graduate Studies, Walailak University, Nakhon Si Thammarat, 80160, Thailand}

\author{Ali \"Ovg\"un}
\email{ali.ovgun@emu.edu.tr}
\affiliation{Physics Department, Eastern Mediterranean University, Famagusta, 99628 North Cyprus via Mersin 10, Turkiye}

\date{\today}

\begin{abstract}

In this work, we investigate wormhole solutions through the utilization of gravitational decoupling, employing the Minimal Geometric Deformation (MGD) procedure within the framework of Trace-Free Gravity. We base our investigation on static and spherically symmetric Morris-Thorne traversable wormholes, considering both constant and variable equation of state parameters. We derive the field equations and extract the shape function for each scenario. Moreover, we explore the gravitational decoupling technique and examine various forms of energy density for both a smeared and particle-like gravitational source, encompassing the realm of noncommutative geometry and a statically charged fluid. We also examinethe wormhole geometry through the utilization of embedding diagrams. Through our analysis, we uncover a violation of the Null Energy Condition (NEC). To conclude, we employ the Gauss-Bonnet theorem to determine the weak deflection angle for the wormhole configurations.

\end{abstract}

\maketitle


\section{Introduction}

A Trace-Free version of the Einstein (TFE) equations provides the resolution of the cosmological constant problem of which the observed cosmological constant is much smaller than the expected value. The formulation was proposed by Weinberg in his review \cite{Weinberg:1988cp}. Indeed, the trace-free gravity is essentially equivalent to adopting unimodular gravity \cite{Finkelstein:2000pg,Henneaux:1989zc,Unruh:1988in,Ng:1990xz,Alvarez:2005iy,Ellis:2010uc,Ellis:2013uxa,Barcelo:2014mua,Burger:2015kie,Alvarez:2015oda,Shaposhnikov:2008xb,Smolin:2010iq,Jain:2012gc,Alvarez:2013fs,Eichhorn:2013xr} and its generalized versions \cite{Barvinsky:2019agh,Barvinsky:2017pmm}. The approach does not only determine a unique value for the effective cosmological constant, but it also does solve the discrepancy between theory and observation in the standard approach. This article is aimed to investigate traversable wormhole solutions in TFE gravity, incorporating the consideration of variable equations of state and taking advantage of gravitational decoupling by means of minimal geometric deformation (MGD) approach.

As widely acknowledged, solving Einstein field equations poses a non-trivial challenge, particularly when dealing with cases invoking spherical symmetry. A recent advancement addressing this complexity is the introduction of a novel methodology known as the gravitational decoupling through the minimal geometric deformation (MGD) scheme \cite{Ovalle:2017fgl,Ovalle:2017wqi}. The Brane-World scenario \cite{Casadio:2012rf,Casadio:2013uma,Ovalle:2013xla,Ovalle:2014uwa,Casadio:2015jva} was originally investigated using such an approach. Subsequently, it evolved into a gravitational source decoupling scheme, facilitating the extension of isotropic spherical solutions of the Einstein field equations to encompass anisotropic domains \cite{Ovalle:2017fgl}. Over time, this approach has been broadly adopted across various branches, see for wormholes \cite{Tello-Ortiz:2021kxg}, playing a pivotal role in expanding or constructing novel solutions for the Einstein equations and their extensions, see e.g., \cite{Casadio:2015gea,Ovalle:2017wqi,Ovalle:2018gic,Tello-Ortiz:2022hyf}.

Gibbons and Werner introduced a novel geometric approach for computing the weak deflection angle, employing the Gauss-Bonnet theorem (GBT) on optical geometries applicable to asymptotically flat spacetimes \cite{Gibbons:2008rj}. This technique involves solving the GBT integral over an infinite domain defined by the light ray boundaries. Subsequently, Werner extended this methodology to stationary spacetimes by incorporating the Finsler-Randers optical geometry, utilizing Nazım’s osculating Riemannian manifolds \cite{Werner:2012rc}.

Building upon Werner's work, Ishihara et al. further extended the method to finite distances, specifically considering scenarios with significant impact parameters, as opposed to relying on asymptotic receiver and source conditions \cite{Ishihara:2016vdc,Ishihara:2016sfv}. T. Ono et al. later applied this finite-distance approach to axisymmetric spacetimes \cite{Ono:2017pie}. Crisnejo and Gallo \cite{Crisnejo:2018uyn} utilized the GBT to derive gravitational light deflections within a plasma medium. More recently, Li et al. investigated the impact of finite distances on the weak deflection angle, introducing massive particles and the Jacobi metric within the framework of GBT \cite{Li:2020dln,Li:2020wvn}. For further developments in this field, one can refer to more recent works on wormholes \cite{Jusufi:2017mav,Ovgun:2018tua,Ovgun:2018fnk,Kumaran:2021rgj} and black holes \cite{Javed:2022kzf,Pantig:2022toh,Okyay:2021nnh,Belhaj:2020rdb,Islam:2020xmy,Kumar:2020hgm,QiQi:2023nex,Huang:2023bto,dePaula:2023ozi,Huang:2022gon,Huang:2022soh,Parbin:2022iwt,Halla:2022xce}.

In this work, we investigate wormhole solutions through the utilization of gravitational decoupling, employing the MGD procedure within the framework of Unimodular Gravity. In Sec.\ref{UN}, we take a recap of the basic concept of Unimodular Gravity and employ the MGD procedure. We then examine static and spherically symmetric Morris-Thorne traversable wormholes in Sec.\ref{wh}. In this section, we consider both constants and variables in the equation of state parameter. We also compute the shape function for each case. In Sec.\ref{dcs}, we consider the gravitational decoupling technique and consider the forms of the energy density for a gravitational source in the context of noncommutative geometry and a statically charged fluid. Here we obtain the decoupled solutions of the wormholes. Moreover, we test the energy conditions. Subsequently, the embedding diagrams of the wormholes are illustrated in Sec.\ref{Emb}. In Sec.\ref{lens}, we use the Gauss-Bonnet theorem to compute the weak deflection angle for wormhole solutions. We conclude our findings in the last section.

\section{Unimodular Gravity \& Its Geometrical Deformation}\label{UN}
A detailed formulation of trace-free Einstein gravity and its relationship with unimodular gravity is given in Refs.\cite{Ellis:2010uc,Ellis:2013uxa}. Recall that in the standard formulation the gravitational field is governed by the Einstein field equations:
\begin{equation}
R_{\mu\nu}-\frac{1}{2}g_{\mu\nu}R +\lambda g_{\mu\nu}=8\pi T^{(m)}_{\mu\nu} \,. \label{eq}
\end{equation}
As a result, in unimodular gravity, the presence of an additional constraint on the metric determinant reduces $10$ independent field equations to $9$ independent field equations. Taking the trace of the field equations (\ref{eq}), we obtain
\begin{equation}
4\lambda= R+8\pi T^{(m)}\,. \label{tr}
\end{equation}
Multiplying each side of Eq.(\ref{tr}) by $\tfrac{1}{4}g_{\mu\nu}$ and adding the result to Eq.(\ref{eq}) yields the
trace-free field equations:
\begin{equation}
G_{\mu\nu}=8\pi {\bar T}_{\mu\nu}\,,
\end{equation}
where we have defined
\begin{equation}
G_{\mu\nu}\equiv R_{\mu\nu}-\frac{1}{4}g_{\mu\nu}R  \,, \label{ghat}
\end{equation}
and 
\begin{equation}
 {\bar T}_{\mu\nu} \equiv \left(T^{(m)}_{\mu\nu} -\frac{1}{4}g_{\mu\nu}T^{(m)}\right)\,.
 \label{That}
\end{equation}

Any extension to the above theory will eventually produce new terms in the effective four-dimensional Einstein equations. These “corrections” are usually handled as part of an effective energy-momentum tensor. In the following, the MGD takes the simplest modification:
\begin{equation}
G_{\mu\nu}=8\pi {\bar T}_{\mu\nu}+\delta ({\rm new\,\, terms})_{\mu\nu}\,,\label{mTFE}
\end{equation}
The new terms in Eq.(\ref{mTFE}) may be viewed as part of an effective energy-momentum tensor, whose explicit form may contain new fields, like scalar, vector, and tensor fields, all of them coming from the new gravitational sector not described by Einstein’s theory. 

Therefore, we have the following modification to the stress-momentum tensor:
\begin{equation}
T_{\mu\nu}={\bar T}_{\mu\nu}+\alpha ({\rm new\,\, terms})_{\mu\nu}=\Big(T^{(m)}_{\mu\nu} -\frac{1}{4}g_{\mu\nu}T^{(m)}\Big)+\delta\, \varepsilon_{\mu\nu}\,.\label{mTFE1}
\end{equation}
It is clear that the case $\delta=0$ yields to the original Unimodular Gravity. We consider
the contribution of two gravitational sources $T^{\mu(m)}_{\nu}$, which is known as the seed source and $\varepsilon^{\mu}_{\nu}$. Here the intensity of influence
of the source $\varepsilon_{\mu\nu}$ over $T^{(m)}_{\mu\nu}$ parametrized by a dimensionaless constant $\delta$. 

\section{Traversable Wormhole Solutions}\label{wh}

We consider a static and spherically symmetric Morris-Thorne traversable wormhole in the Schwarzschild coordinates $(t, r, \theta, \phi)$ given by [4]
\begin{equation}
ds^2 = -e^{2\Phi(r)}dt^2  + \frac{1}{\left(1-\frac{b(r)}{r}\right)} dr^2 + r^2 \left(d\theta^2 +\sin^2 \theta d\phi^2\right)\,, \label{met}
\end{equation}
where $\Phi(r)$ and $b(r)$ are the redshift and shape functions, respectively. They are functions of the radial coordinate $r$ only. In the wormhole geometry, the redshift function $\Phi$ should be finite in order to avoid the formation of an event horizon. The radial coordinate $r$ ranges from a minimum value $r_0$, corresponding to the throat of the wormhole, where $b(r_0) = r_0$ at $r = r_0$. A crucial aspect of wormholes is the flaring-out condition, expressed as $b(r) - b'(r) r \geq 0$ in the vicinity of the throat, where a prime denotes a derivative with respect to the radial coordinate $r$. Additionally, it is required that $b(r)/r \rightarrow 0$ as $r \rightarrow \infty$. It is worth noting that the supplementary condition $b(r)/r < 1$ is also enforced. We define a perfect fluid source with energy-momentum tensor as
\beq
{\bar T}_{ab} = ({\bar \rho} + {\bar p}_{t})U_a U_b + {\bar p}_{t} g_{ab}+({\bar p}_{r}-{\bar p}_{t})X_{a}X_{b} \,,\label{smt}
\eeq
where $\bar{\rho}$ is the energy density measured by a comoving observer with the fluid, and $U_{a}$ and $X_{a}$ are its timelike four-velocity and a spacelike unit vector orthogonal to $U_{a}$ and angular directions, respectively. We define an appropriate frame of the fluid velocity vectors \cite{Cadoni:2020izk}
\beq
U^{a}=  e^{-\Phi}\delta^{a}_{0}\,, \quad X^{a}= \sqrt{1-\frac{b(r)}{r}} \delta^{a}_{1}\,,
\eeq
so that $U_{a}U^{a}=-1$ and $X_{a}X^{a}=1$. Here we are working in geometrized units setting the gravitational constant $G$ and the speed of light $c$ to unity. 
The trace-free components of the energy-momentum tensor Eq.(\ref{smt}) in this case read
\beq
{\bar T}_{ab} &=&  \biggl(\frac{1}{4} e^{2 \Phi} \left({\bar p}_r + 2 {\bar p}_t + 3 {\bar \rho} \right),
    \frac{1}{4}(3 {\bar p}_r - 2 {\bar p}_t + {\bar \rho})
    \frac{1}{(1-\frac{b}{r})},\nonumber\\&&\quad\quad
    \frac{1}{4} r^2 \left(-{\bar p}_r + 2 {\bar p}_t +{\bar \rho}\right),
    \frac{1}{4} r^2 \sin^2(\theta) \left(-{\bar p}_r + 2 {\bar p}_t +{\bar \rho} \right)\biggl). \label{3}
\eeq
For $(t,t)$ component, we find
\begin{align}
{\cal G}_{tt} = \frac{e^{2 \Phi }}{4 r^2}(b' (2-r \Phi ')+2 r(2 \Phi '+r \Phi '^2+r \Phi '')-b (3 \Phi '+2 r \Phi '^2+2 r \Phi '')),
\end{align}
while for $(r,r)$ component,
\begin{align}
{\cal G}_{rr}= \frac{1}{4 (b-r) r^2}(b (4+5 r \Phi '-2 r^2 \Phi '^2-2 r^2 \Phi '')+r (-b'(2+r \Phi ')+2 r (-2 \Phi '+r \Phi '^2+r \Phi ''))),
\end{align}
and for $(\theta, \theta)$ component,
\begin{align}
{\cal G}_{\theta\theta}=\frac{1}{4 r}(b (2+r \Phi '-2 r^2 \Phi '^2-2 r^2 \Phi '')+r^2 (-b' \Phi '+2 r (\Phi'^2+\Phi ''))),
\end{align}
where ${\cal G}_{\theta\theta}=\sin^{2}\theta {\cal G}_{\phi\phi}$. Then using the information of metric (\ref{ghat}) and (\ref{That}) in the TFE for the general form, we come up with
\beq
8\pi \left[\left({\bar p}_r + 2 {\bar p}_t + 3 {\bar \rho}\right)\right]&=& \frac{1 }{ r^2}(b' (2-r \Phi ')+2 r(2 \Phi '+r \Phi '^2+r \Phi '')\nonumber\\&&-b (3 \Phi '+2 r \Phi '^2+2 r \Phi ''))\,,\\
8\pi \left[(3 {\bar p}_r - 2 {\bar p}_t + {\bar \rho})\right] &=&\frac{-1}{ r^3}(b (4+5 r \Phi '-2 r^2 \Phi '^2-2 r^2 \Phi '')+r (-b'(2+r \Phi ')\nonumber\\&&+2 r (-2 \Phi '+r \Phi '^2+r \Phi '')))\,,\\
8\pi \left[\left(-{\bar p}_r + 2 {\bar p}_t +{\bar \rho}\right)\right]&=&\frac{1}{ r^3}(b (2+r \Phi '-2 r^2 \Phi '^2-2 r^2 \Phi '')\nonumber\\&&+r^2 (-b' \Phi '+2 r (\Phi
'^2+\Phi '')))\,.
\eeq
We can directly derive the Einstein field equations for $\Phi'=0$ to obtain
\beq
\frac{b'}{4\pi r^2} &=&{\bar p}_r + 2 {\bar p}_t + 3 {\bar \rho} \equiv {\bar \rho}^{\rm eff.}\,,\label{eq1}\\
- \frac{2 b-rb'}{4\pi r^3}&=& 3 {\bar p}_r - 2 {\bar p}_t + {\bar \rho} \equiv {\bar p}^{\rm eff.}_{r}\,,\label{eq2}\\
\frac{b}{4\pi r^3}&=& -{\bar p}_r + 2 {\bar p}_t +{\bar \rho} \equiv {\bar p}^{\rm eff.}_{t}\,.\label{eq3}
\eeq

Therefore, the effective matter sector of the present consideration is given by
\beq
\frac{2 b'}{r^2} &=&8\pi \left[\left({\bar p}_r + 2 {\bar p}_t + 3 {\bar \rho} \right)+\delta\varepsilon^{0}_{0}\right]\equiv 8\pi \rho^{\rm eff.},\label{nde}\,\\
- \frac{4 b-2rb'}{r^3}&=&8\pi \left[(3 {\bar p}_r - 2 {\bar p}_t + {\bar \rho} )-\delta\varepsilon^{1}_{1}\right]\equiv 8\pi p^{\rm eff.}_{r}\,,\label{nde1}\\
\frac{2 b}{r^3}&=&8\pi \left[\left(-{\bar p}_r + 2 {\bar p}_t +{\bar \rho} \right)-\delta\varepsilon^{2}_{2}\right]\equiv 8\pi p^{\rm eff.}_{t}\label{nde2}\,.
\eeq
To split the complex set of Eqs. (\ref{nde})–(\ref{nde2}), we implement the gravitational decoupling by means of the MGD. In this case, the minimally deformed shape function $b(r)$ is given by
\beq
b(r) \rightarrow b(r)+\delta f(r)\,,\label{dec}
\eeq
with ${\bar b}(r)$ being the original shape function given in the preceding section and $f(r)$ the decoupler function. Basically, values of $\delta$ could be small. Putting Eq.(\ref{dec}) into the set (\ref{nde})–(\ref{nde2}), we obtain the following system of equations
\beq
\frac{2 b'}{r^2} &=&8\pi \left[\left({\bar p}_r + 2 {\bar p}_t + 3 {\bar \rho} \right)\right]\,,\\
- \frac{4 b-2r b'}{r^3}&=&8\pi \left[(3 {\bar p}_r - 2 {\bar p}_t + {\bar \rho} )\right]\,,\\
\frac{2 b}{r^3}&=&8\pi \left[\left(-{\bar p}_r + 2 {\bar p}_t +{\bar \rho} \right)\right]\,.
\eeq
The second set of equations is given by
\beq
\frac{2f'(r)}{r^{2}} &=&8\pi \varepsilon^{0}_{0}\,,\label{de1}\\  \frac{4 f(r)-2r f'(r)}{r^3} &=&8\pi \varepsilon^{1}_{1}\,,\label{de2}\\
-\frac{2 f(r)}{r^3}&=&8\pi \varepsilon^{2}_{2}\,.\label{de3}
\eeq
To obtain specific forms of $f(r)$, we will consider two types of density profiles generated in noncommutative geometry and a statically charged fluid. 

\subsection{Constants $\omega$ \& $\beta$}

\begin{figure}[ht!]
    \centering
    \includegraphics[width=0.48\textwidth]{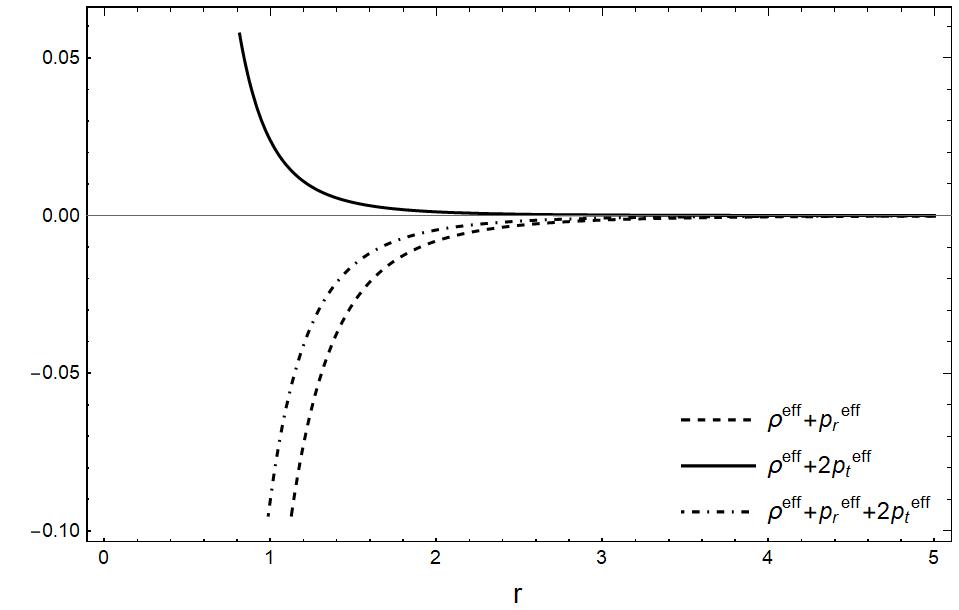}
    \caption{We show the behaviors of $\rho^{\text{eff}} +  p_r^{\text{eff}} $, $\rho^{\text{eff}} + 2 p_t^{\text{eff}} $ and $\rho^{\text{eff}}  + p_r^{\text{eff}} +2p_t^{\text{eff}}  $ versus $r$ using $\beta=-1.1$,  $r_0 =1$ and $\omega =0.8 $. }
    \label{weako}
\end{figure}

Basically, it is first common to assume the barotropic equations of state given as ${\bar p}_r=\omega {\bar \rho}$ with $\omega$ being a constant as well as $p_t=\beta p_{r}$ with constant $\beta$. As mentioned in Ref.\cite{Agrawal:2022atn}, the values of $\omega$ can be constrained to $-1.5 \leq \omega \leq 1$. This allows us to describe various types of cosmological fluids, for example stiff matter \cite{Carr:2010wk}, radiation \cite{Bahamonde:2016ixz}, Dustlike \cite{Kashargin:2020miw}, dark energy \cite{Wang:2016pwd}, phantom fluid \cite{Cataldo:2013ala}, and holographic dark energy \cite{Garattini:2023wgk}. Moreover, we also need $\beta\neq 1$ to guarantee anisotropy and $\beta\neq 0$ to avoid singularities. From Eq.(\ref{eq1})-Eq.(\ref{eq3}), we find
\beq
- \frac{b}{8\pi r^3} + \frac{rb'}{8\pi r^3} &=&{\bar p}_r + {\bar \rho} = (1+\omega){\bar \rho} \,,\label{eq12}\\
\frac{r (\omega -1) b'(r)+(\omega +3) b(r)}{16 \pi (\omega +1) r^3}&=& {\bar p}_t \,.\label{eq13}
\eeq
We can simply solve for the analytical solution of the above system to obtain
\beq
b(r) = r_{0}\,\Big(\frac{r}{r_{0}}\Big)^{\frac{2 \omega \beta +\omega +3}{2 \omega \beta -\omega +1}}\,,\label{to}
\eeq
\begin{figure}[ht!]
    \centering
\includegraphics[width=3in,height=3in,keepaspectratio=true]{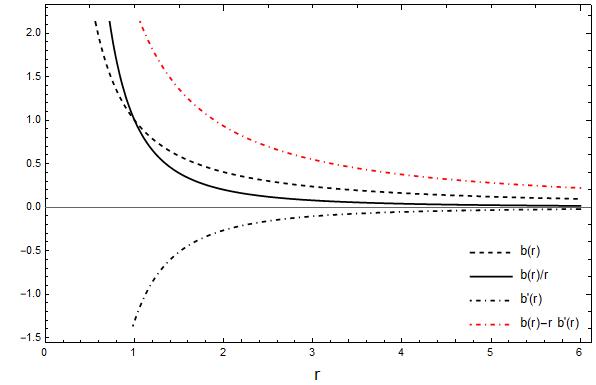}
\includegraphics[width=3in,height=3in,keepaspectratio=true]{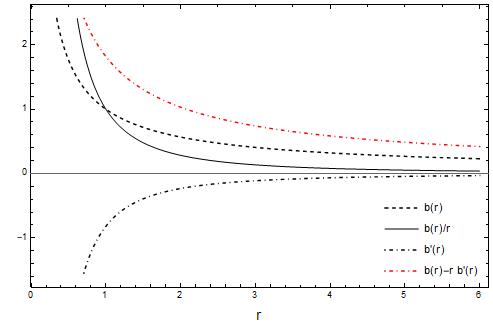}
    \caption{We show the behaviors of $b(r), b'(r), b(r)/r$ and $b(r)-r b'(r)$ versus $r$ in Eq.(\ref{to}) for $\beta=-1.1$ and $r_0 =1$ and $\omega = 0.8$ (Left Panel) and $\omega = 1$ (Right Panel).}
    \label{fig:enter-label0on}
\end{figure}
matching with that found in Ref.\cite{Agrawal:2022atn}. We recover the Ellis-Bronnikov (EB) shape function, $b(r)=\tfrac{r^{2}_{0}}{r}$, when $\omega=-\beta^{-1}$. We can show that the shape function (\ref{to}) satisfies the usual properties displayed in Fig.(\ref{fig:enter-label0on}). We can also simply check the expression of the flare-out condition given by
\beq
\frac{b(r)-r b'(r)}{b(r)^{2}} = -\frac{2 (\omega +1) \left(\frac{r}{r_{0}}\right)^{-\frac{2 \beta  \omega +\omega +3}{2 \beta  \omega -\omega +1}}}{(2 \beta -1) r_{0} \omega +r_{0}} {\underset{r=r_{0}}{=}} -\frac{2 (\omega +1)}{(2 \beta -1) r_{0} \omega +r_{0}} >0\,.\label{newb01}
\eeq
which yields the constraints on $\omega$ and $\beta$:
\beq
\omega >0\land \beta <\frac{\omega -1}{2 \omega }\,.
\eeq
In this case, we can easily compute the energy density to obtain
\beq
{\bar \rho}(r) = -\frac{1}{8 \pi  (\omega +1) r^2}\left(\frac{r}{r_{0}}\right)^{\frac{2 (\omega +1)}{\omega (2 \beta -1)+1}}\,.\label{newro1}
\eeq
which, on the throat becomes
\beq
\rho(r_{0}) = -\frac{1}{8 \pi  (\omega +1) r_{0}^2}\,.\label{newb301}
\eeq
Using $\rho(r)$, we can compute $p_{r}$ and $p_{t}$ to obtain
\beq
{\bar p}_{r} &=& -\frac{\omega}{8 \pi  (\omega +1) r^2}\left(\frac{r}{r_{0}}\right)^{\frac{2 (\omega +1)}{\omega (2 \beta -1)+1}}\,,\label{inpr}\\ {\bar p}_t &=&-\frac{\omega\beta}{8 \pi  (\omega +1) r^2}\left(\frac{r}{r_{0}}\right)^{\frac{2 (\omega +1)}{\omega  (2 \beta -1)+1}}\,.
\eeq
We see from Fig.(\ref{weako}) for constants $\omega$ \& $\beta$, the NEC and SEC are violated at the wormhole throat $r = r_0$. In other words, we
discover that at the wormhole throat $r = r_0$, we have for the energy condition $\rho^{\rm eff.} + p^{\rm eff.}_{r} < 0$, along with the condition $\rho^{\rm eff.} + p^{\rm eff.}_{r} + 2p^{\rm eff.}_{t} < 0$, by arbitrary small values.

\subsection{Variable $\omega$ \& Constant $\beta$}
In the preceding subsection, we have worked on the constant EoS and discussed some important properties of the shape function $b(r)$ illustrated in Fig.(\ref{fig:enter-label0on}). However, variable EoS can also be considered. In case of $\omega=\omega(r)$, we come up with
\beq
\int\frac{1}{b(r)}\,d b(r)=\int\frac{(2 \beta  \omega (r)+\omega (r)+3)}{r (2 \beta  \omega (r)-\omega (r)+1)}\,dr\,.\label{newbo}
\eeq
Therefore, we obtain
\beq
b(r)=r_{0}\,\exp\Big(\int\frac{(2 \beta  \omega (r)+\omega (r)+3)}{r (2 \beta  \omega (r)-\omega (r)+1)}\,dr\Big)\,.\label{newbomgg}
\eeq
with $r_{0}$ being a constant. Let us take a more general EoS:
\beq
\omega(r)=\frac{1}{\mu r}\,.\label{mu}
\eeq
We find that
\beq
\omega(r){\underset{r=r_{0}}{\rightarrow}}\frac{1}{\mu r_{0}},\,\,\,\omega(r){\underset{\mu\rightarrow 0}{\rightarrow}}\infty\,,\quad{\rm and}\quad \omega(r){\underset{r\,{\rm or}\,\mu\rightarrow \infty}{\rightarrow}}0\,.
\eeq
We solve Eq.(\ref{newbomgg}) to obtain
\beq
b(r)=r_{0}\,\left(\frac{r}{r_{0}}\right)^{\frac{2 \beta +1}{2 \beta -1}}\left(\frac{1-2 \beta -\mu  r}{1-2 \beta -\mu  r_{0}}\right)^{\frac{4 (\beta -1)}{2 \beta -1}}\,.\label{newbmu}
\eeq
We can also simply check the expression of the flare-out condition given by
\beq
\frac{b(r)-r b'(r)}{b(r)^{2}}\Big|_{r=r_{0}} = \frac{2 \mu  r_{0}+2}{-\mu r_{0}^2-2 \beta  r_{0}+r_{0}}>0\,.\label{newbn}
\eeq
For more convenience, we take instead $\mu=\tfrac{\mu_{1}}{r_{0}}$ to have
\beq
\frac{b(r)-r b'(r)}{b(r)^{2}}\Big|_{r=r_{0}} = \frac{2 \mu_{1}+2}{-2 \beta  r_{0}-\mu_{1} r_{0}+ r_{0}}>0\,.\label{newbn1}
\eeq
With a flare-out condition, we find the constraints on $\beta$ and $\mu_{1}$:
\beq
\beta \leq 1\land -1< \mu_{1}<1-2 \beta \land r_{0}>0\,.
\eeq
If we take $\beta=-1.1$, we find that $-1 < \mu_{1} < 3.2$. However, in the following, we consider $\mu=-\tfrac{\beta}{r_{0}}$ and write
\beq
\omega(r)=-\frac{r_0}{\beta r}\,,\label{omg}
\eeq
where
\beq
\omega(r){\underset{r=r_{0}}{\rightarrow}}-\frac{1}{\beta}\,,\quad{\rm and}\quad \omega(r){\underset{r\rightarrow \infty}{\rightarrow}}0\,.
\eeq
In this case, $\beta$ is the only free parameter. Substituting Eq.(\ref{omg}) into Eq.(\ref{newbomgg}), we find
\beq
b(r)=r_{0}\left(\frac{r}{r_{0}}\right)^{\frac{2 \beta +1}{2 \beta -1}}\left(\frac{\beta  r-2 \beta  r_{0}+r_{0}}{r_{0}-\beta r_{0}}\right)^{\frac{4 (\beta -1)}{2 \beta -1}}\,. \label{eqn46}
\eeq
\begin{figure}[ht!]
    \centering
\includegraphics[width=3in,height=3in,keepaspectratio=true]{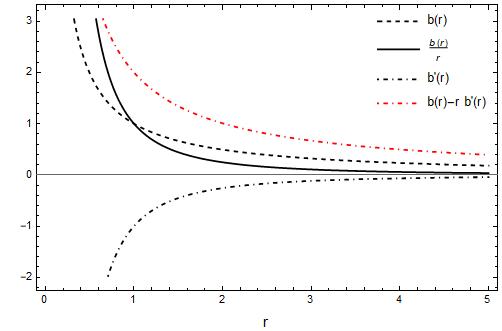}
\includegraphics[width=3in,height=3in,keepaspectratio=true]{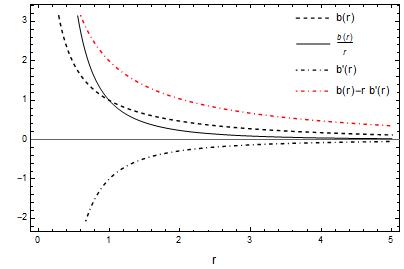}
    \caption{We show the behaviors of $b(r), b'(r), b(r)/r$ and $b(r)-r b'(r)$ versus $r$ in Eq.(\ref{eqn46}) for  $r_0 =1$ and $\beta=-0.01$  (Left Panel) and  $\beta=-0.05$ (Right Panel).}
    \label{fig:enter-label0not}
\end{figure}
We can quantify the features of a shape function $b(r)$. Assuming that $|\beta| \ll {\cal O}(1)$, we can expand $b(r)$ to the first order of $\beta$ to obtain
\beq
b(r) = \frac{r_{0}^2}{r}+\frac{4}{r}\left(-r_{0}^2 \log \left(\frac{r}{r_{0}}\right)+r r_{0}-r_{0}^2\right)\beta+{\cal O}(\beta^{2})\,.\label{expand}
\eeq
We see that when $|\beta| \ll {\cal O}(1)$ a shape function satisfies the usual properties as it should be. Notice that we can obtain the EB shape function, $b(r)=\tfrac{r^{2}_{0}}{r}$, when $\beta=0$. The behaviors of $b(r)$ can be seen in Fig.(\ref{fig:enter-label0not}). In this case, we can easily compute the energy density to obtain
\beq
{\bar \rho}(r) = -\frac{\beta }{4 \pi  (\beta -1) r^2}\left(\frac{r}{r_{0}}\right)^{\frac{2}{2 \beta -1}+1} \left(\frac{\beta  r-2 \beta  r_{0}+r_{0}}{r_{0}-\beta r_{0}}\right)^{\frac{2}{1-2 \beta }+1}\,.\label{newromg}
\eeq
which, on the throat becomes
\beq
{\bar \rho}(r_{0}) = -\frac{\beta }{4 \pi  (\beta -1) r_{0}^2}\,.\label{newb301omg}
\eeq
Using ${\bar \rho}(r)$, we can compute ${\bar p}_{r}$ and ${\bar p}_{t}$ to obtain
\begin{figure}[ht!]
    \centering
    \includegraphics[width=0.48\textwidth]{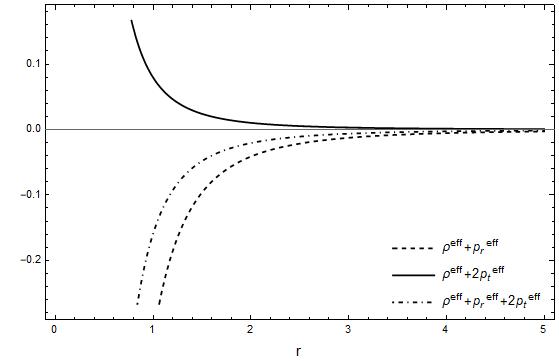}
    \caption{We show the behaviors of $\rho^{\text{eff}} +  p_r^{\text{eff}} $, $\rho^{\text{eff}} + 2 p_t^{\text{eff}} $ and $\rho^{\text{eff}}  + p_r^{\text{eff}} +2p_t^{\text{eff}}  $ versus $r$ using $\beta=-0.01$, and $r_0 =1$. }
    \label{chargedcon}
\end{figure}
\beq
{\bar p}_{r} &=& \frac{1}{4 \pi  (\beta -1) r^2}\left(\frac{r}{r_{0}}\right)^{\frac{2}{2 \beta -1}} \left(\frac{\beta  r-2 \beta r_{0}+r_{0}}{r_{0}-\beta r_{0}}\right)^{\frac{2}{1-2 \beta }+1}\,,\label{inpromg}\\ {\bar p}_t &=&\frac{\beta}{4 \pi  (\beta -1) r^2}\left(\frac{r}{r_{0}}\right)^{\frac{2}{2 \beta -1}} \left(\frac{\beta  r-2 \beta r_{0}+r_{0}}{r_{0}-\beta r_{0}}\right)^{\frac{2}{1-2 \beta }+1}\,.
\eeq
In this case, we also discover that the NEC \& SEC are violated, i.e., $\rho^{\text{eff}} +  p_r^{\text{eff}}<0,\,\rho^{\text{eff}}  + p_r^{\text{eff}} +2p_t^{\text{eff}}  <0$, by arbitrary small values, see Fig.(\ref{chargedcon}).

\section{Decoupled solutions}\label{dcs}
We consider the gravitational decoupling technique and consider two forms of the energy
density for a smeared and particle-like gravitational source in the context of noncommutative geometry and a statically charged fluid.

\subsection{Non-commutative Geometry Density Profiles}\label{den}
In the context of noncommutative geometry, an interesting development of string/M-theory involves the requirement for quantizing spacetime. The non-commutativity of spacetime is encoded in the commutator $[{\bf x}^{\mu},{\bf x}^{\nu}]=i \theta_{\mu\nu}$, where $\theta_{\mu\nu}$ is an antisymmetric matrix which determines the fundamental discretization of spacetime. It has also been shown that noncommutativity flavors the smeared objects instead of the point-like structures in flat spacetime \cite{Smailagic:2003yb}. Mathematically, the smearing permits substitution of the Dirac-delta function by a Gaussian distribution of minimal length $\sqrt{\theta}$.

Specifically, the formulation of the energy density for a gravitational source that is static, spherically symmetric, and resembles both a smeared and particle-like structure has been examined \cite{Nicolini:2005vd,Lobo:2010uc} given by
\beq
\rho_{\theta}(r) = \frac{M}{(4\pi\theta)^{3/2}}\exp\Big(-\frac{r^{2}}{4\theta}\Big)\,,\label{newb30}
\eeq
where the mass $M$ is diffused throughout a region of linear size $\sqrt{\theta}$ due to the intrinsic uncertainty encoded in the coordinate commutator. Like black holes, see e.g., Refs.\cite{Nicolini:2005vd,DiGrezia:2007bxw}, the wormhole metric is expected to be modified when a noncommutative spacetime is taken into account, see \cite{Rahaman:2014dpa,Garattini:2008xz,Rahman:2022pug}. Moreover, we also get inspired by the work of Mehdipour when searching for a new fluid model. A Lorentzian distribution of particle-like gravitational source permits possible energy density profile as given in Ref.\cite{Mehdipour:2011mc,Rahaman:2014dpa,Liang:2012vx} as follows:
\beq
\rho_{\phi}(r) = \frac{M\sqrt{\phi}}{\pi^{2}(r^{2}+\phi)^{2}}\,\label{newb3011}
\eeq
with $\phi$ being the noncommutativity parameter. We next consider Eqs. (\ref{de1})-(\ref{de3}) and employ the density profiles given above to quantify the decoupled solutions, $f(r)$. We assume that $\varepsilon^{0}_{0}=\rho_{\theta,\phi}$ and solve for $f(r)$ to obtain.
\beq
f(r) = \begin{cases}
  M \left(\text{erf}\left(\frac{r}{2 \sqrt{\theta }}\right)-\text{erf}\left(\frac{r_{0}}{2 \sqrt{\theta }}\right)-\frac{r}{\sqrt{\pi } \sqrt{\theta }}e^{-\frac{r^2}{4 \theta }}+\frac{r_{0}}{\sqrt{\pi } \sqrt{\theta }}e^{-\frac{r_{0}^2}{4 \theta }}\right)  & \text{for}\,\rho_{\theta}=\varepsilon^{0}_{0}\,, \\
  \frac{2 M}{\pi }\left( \frac{r_{0}/\sqrt{\phi }}{\frac{r_{0}^2}{\phi}+1}-\frac{r/\sqrt{\phi }}{\frac{r^2}{\phi}+1 }+\tan ^{-1}\left(\frac{r}{\sqrt{\phi }}\right)-\tan ^{-1}\left(\frac{r_{0}}{\sqrt{\phi }}\right)\right) & \text{for}\,\,\rho_{\phi}=\varepsilon^{0}_{0}\,.
\end{cases}
\eeq

We can simply check that a condition of $f(r=r_0)=0$ is satisfied. Substituting $f(r)$ to Eq.(\ref{to}), the original (traversable) wormhole solutions can be geometrically deformed. In this case, the original wormhole solution will be deformed by the above results. Therefore, we obtain for $\rho_{\theta}=\varepsilon^{0}_{0}$
\beq
\frac{b(r)}{\sqrt{\theta }} &=&   \frac{r_{0}}{\sqrt{\theta }}\,\Big(\frac{r/\sqrt{\theta }}{r_{0}/\sqrt{\theta }}\Big)^{\frac{2 \omega  \beta +\omega +3}{2 \omega  \beta -\omega +1}}\nonumber\\&+&\delta \frac{M}{\sqrt{\theta }}\left(\text{erf}\left(\frac{r}{2 \sqrt{\theta }}\right)-\text{erf}\left(\frac{r_{0}}{2 \sqrt{\theta }}\right)-\frac{r}{\sqrt{\pi } \sqrt{\theta }}e^{-\frac{r^2}{4 \theta }}+\frac{r_{0}}{\sqrt{\pi } \sqrt{\theta }}e^{-\frac{r_{0}^2}{4 \theta }}\right) \,,\label{bc1}
\eeq
and for $\rho_{\phi}=\varepsilon^{0}_{0}$  
\beq
 \frac{b(r)}{\sqrt{\phi}} &=&  \frac{r_{0}}{\sqrt{\phi}}\,\Big(\frac{r/\sqrt{\phi}}{r_{0}/\sqrt{\phi}}\Big)^{\frac{2 \omega  \beta +\omega +3}{2 \omega  \beta -\omega +1}}\nonumber\\&+&\frac{2\delta}{\pi }\frac{M}{\sqrt{\phi}}\left( \frac{r_{0}/\sqrt{\phi }}{\frac{r_{0}^2}{\phi}+1}-\frac{r/\sqrt{\phi }}{\frac{r^2}{\phi}+1 }+\tan ^{-1}\left(\frac{r}{\sqrt{\phi }}\right)-\tan ^{-1}\left(\frac{r_{0}}{\sqrt{\phi }}\right)\right)\,.\label{bc2}
\eeq
\begin{figure}[ht!]
    \centering
\includegraphics[width=3in,height=3in,keepaspectratio=true]{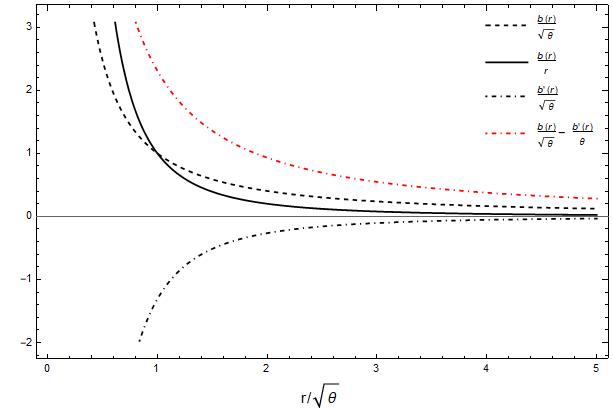 }
\includegraphics[width=3in,height=3in,keepaspectratio=true]{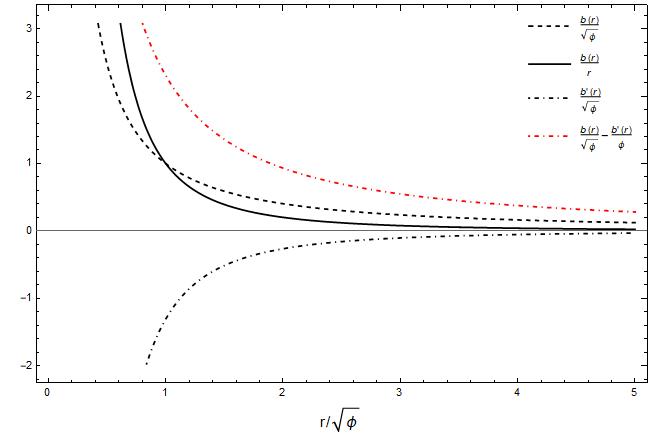 }
    \caption{We show the behaviors of $\frac{b(r)}{\sqrt{\theta}}, \frac{b'(r)}{\sqrt{\theta}}, \frac{b(r)}{\sqrt{\theta}}r$ and $\frac{b(r)-r b'(r)}{\sqrt{\theta}}$ versus $\frac{r}{\sqrt{\theta}}$ in Eq.(\ref{bc1}) for $\beta=-1.1$, $\omega=0.8$, $r_0 =1$,  $\delta = 0.001$ and $\frac{M}{\sqrt{\theta}}=1$ (Left Panel) and $\frac{b(r)}{\sqrt{\phi}}, \frac{b'(r)}{\sqrt{\phi}}, \frac{b(r)}{\sqrt{\phi}}r$ and $\frac{b(r)-r b'(r)}{\sqrt{\phi}}$ versus $\frac{r}{\sqrt{\phi}}$ in Eq.(\ref{bc2}) for $\beta=-1.1$, $\omega=0.8$, $r_0 =1$,  $\delta = 0.001$ and $\frac{M}{\sqrt{\phi}}=1$ (Right Panel).}
    \label{fig:enter-label}
\end{figure}
We can show that the shape function (\ref{bc1}) and (\ref{bc2}) satisfy the usual properties, see Fig.(\ref{fig:enter-label}), namely the throat
condition and so on. More specifically, using $b'(r_{0})<1$, we have for Eq.(\ref{bc1}):
\beq
\delta <\frac{1}{\frac{M}{\sqrt{\theta}}r_{0}^{2}}\Big(2 \sqrt{\pi } m e^{\frac{r_{0}^2}{4}}+2 \sqrt{\pi } e^{\frac{r_{0}^2}{4}}\Big)\,,
\eeq
and for Eq.(\ref{bc2}):
\beq
\delta <\frac{e^{\frac{r_{0}^2}{4}}8 \sqrt{\pi }}{\frac{M}{\sqrt{\phi}}r_{0}^2(m r_{0}-\beta+2)}\Big(m r_{0}+1 \Big)\,,
\eeq
where $m=\frac{2 \omega  \beta +\omega +3}{2 \omega  \beta -\omega +1}$.

\subsection{Electromagnetic field}\label{EM}
\begin{figure}[ht!]
    \centering
\includegraphics[width=3in,height=3in,keepaspectratio=true]{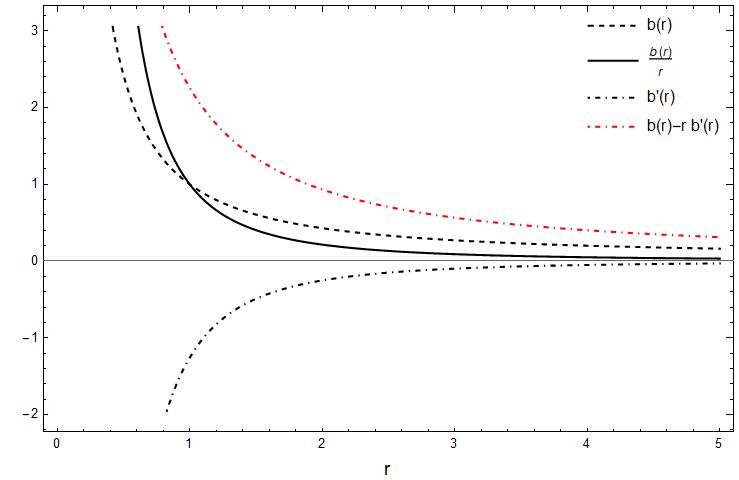}
\includegraphics[width=3in,height=3in,keepaspectratio=true]{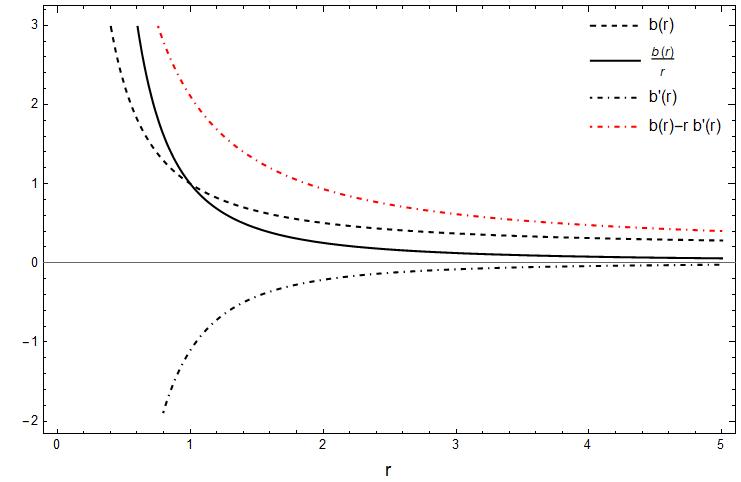}
    \caption{We show the behaviors of $b(r), b'(r), b(r)/r$ and $b(r)-r b'(r)$ versus $r$ in Eq.(\ref{newomg}) for $\beta=-1.1$, $r_0 =1$ and ${\cal Q} =1$ (Left Panel) and ${\cal Q}=2$ (Right Panel).}
    \label{fig:enter-label0em}
\end{figure}

The procedure used to obtain wormhole solutions can be extended to include also the electromagnetic field as an additional source. Here we include the contribution of an electric field generated by a point charge, $\cal{Q}$. The Einstein–Maxwell equations for a statically charged fluid can be given in terms of density $\rho(r)$, radial pressure $p_r$, and tangential pressure $p_t$. We follow the algebraic structure of stress-energy tensors for electromagnetic fields given in \cite{Dymnikova:2004zc,Dymnikova:2021dqq} by $T^{0}_{0}=T^{1}_{1}$ implying that $\rho=-p_{r}$. A pure spherically symmetric electromagnetic field without the contribution of any additional sources is determined by
\beq
T^{\rm EM}_{\mu\nu}=\frac{{\cal Q}^{2}}{8\pi r^{4}}{\rm diag.}(1,-1,1,1)\,,
\eeq
which is conserved and traceless. We assume that $\varepsilon^{0}_{0}=T^{0}_{0}$ and solve for $f(r)$ to obtain.
\beq
f(r) = \frac{{\cal Q}^2}{2}\Big(\frac{1}{r_{0}}-\frac{1}{r}\Big)\,,
\eeq
which a condition of $f(r=r_0)=0$ is assumed. In this case, the original wormhole solution will be deformed by the above result. Therefore, we obtain
\beq
b(r) =   r_{0}\,\Big(\frac{r}{r_{0}}\Big)^{\frac{2 \omega \beta +\omega +3}{2 \omega \beta -\omega +1}}+\delta \frac{{\cal Q}^2}{2}\Big(\frac{1}{r_{0}}-\frac{1}{r}\Big) \,.\label{bem}
\eeq
We can show that the shape function (\ref{bc1}) and (\ref{bc2}) satisfy the usual properties, see Fig.(\ref{fig:enter-label0em}), namely the throat
condition and so on. We can also simply check the expression of the flare-out condition given by
\beq
\frac{b(r)-r b'(r)}{b(r)^{2}} = \frac{2 r r_{0} \left(\delta  {\cal Q}^2 ((2 \beta -1) \omega +1) (r-2 r_{0})-4 r r_{0}^2 (\omega +1) \left(\frac{r}{r_{0}}\right)^{\frac{2 \beta  \omega +\omega +3}{2 \beta  \omega -\omega +1}}\right)}{((2 \beta -1) \omega +1) \left(\delta  {\cal Q}^2 r-\delta  {\cal Q}^2 r_{0}+2 r r_{0}^2 \left(\frac{r}{r_{0}}\right)^{\frac{2 \beta  \omega +\omega +3}{2 \beta  \omega -\omega +1}}\right)^2}\,,\label{newb01}
\eeq
which yields the constraints at $r=r_{0}$:
\beq
\frac{\delta  {\cal Q}^2 (-2 \beta  \omega +\omega -1)-4 r_{0}^2 (\omega +1)}{2 r_{0}^3 ((2 \beta -1) \omega +1)}>0\quad{\longrightarrow}\quad {\cal Q}^{2}<\frac{-4 r_{0}^2 \omega -4 r_{0}^2}{2 \beta  \delta  \omega -\delta  \omega +\delta }\,,\label{conem}
\eeq
with the following conditions:
\beq
\delta >0\land \omega >0\land \left(r_{0}>0\land \beta <\frac{\omega -1}{2 \omega }\right)\,.
\eeq
Moreover, in case of $\omega=\omega(r)$ and constant $\beta$, we have
\beq
b(r)=r_{0}\left(\frac{r}{r_{0}}\right)^{\frac{2 \beta +1}{2 \beta -1}}\left(\frac{\beta  r-2 \beta  r_{0}+r_{0}}{r_{0}-\beta r_{0}}\right)^{\frac{4 (\beta -1)}{2 \beta -1}}+\delta \frac{{\cal Q}^2}{2}\Big(\frac{1}{r_{0}}-\frac{1}{r}\Big)\,.\label{newomg}
\eeq 
We find for the flare-out condition to be satisfied:
\beq
\delta >0\land r_{0}>0\land |{\cal Q}|<2 \sqrt{\frac{r_{0}^2}{\delta }}\,.
\eeq
Notice that values of ${\cal Q}$ depend on a parameter $\delta$ and do not depend on $\beta$ in this case.

\section{Embedding diagram}\label{Emb}
\begin{figure}[ht!]
    \centering
\includegraphics[width=4in,height=4in,keepaspectratio=true]{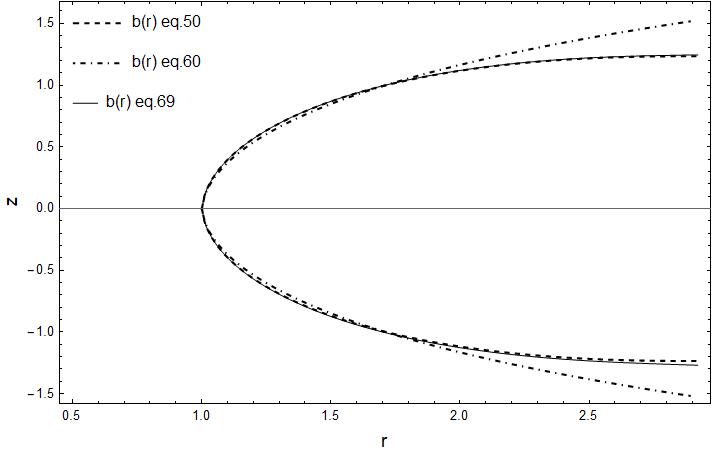}
    \caption{We display the embedding diagrams of various shape functions given in Eq.(\ref{eqn46}), (\ref{bc2}), and (\ref{newomg}).}
    \label{Embedding}
\end{figure}

In this section, we analyse the embedding diagrams to represent the wormhole solutions by considering an equatorial slice $\theta=\pi/2$ at some fix moment in time $t$ = constant. The metric then becomes
\begin{equation}
ds^2 =  \frac{1}{\left(1-\frac{b(r)}{r}\right)} dr^2 + r^2 d\phi^2\,, \label{emb1}
\end{equation}
Having embed the metric (\ref{emb1}) into three-dimensional Euclidean space, we can visualize this slice. Here we parameterize spacetime using the cylindrical coordinates as
\begin{equation}
ds^2 =  dz^{2}+dr^{2}+ r^2 d\phi^2\,, \label{emb1c}
\end{equation}
which can be rewritten as
\begin{equation}
ds^2 =  \Big(1+\Big(\frac{dz}{dr}\Big)^{2}\Big)dr^{2}+ r^2 d\phi^2\,. \label{emb1c1}
\end{equation}
Having compared Eq.(\ref{emb1}) with Eq.(\ref{emb1c1}), we come up with:
\begin{equation}
\frac{dz}{dr} = \pm \sqrt{\Big(\frac{r}{r-b(r)}-1\Big)} \,. \label{emb1c1s}
\end{equation}
To test the results, we consider $b(r)$ given by Eq.(\ref{bc2}) and Eq.(\ref{newomg}). Invoking numerical techniques allows us to illustrate the wormhole shape given in Fig.(\ref{Embedding}). In Fig.(\ref{Embedding}), we consider various shape functions $b(r)$ given in Eq.(\ref{eqn46}), (\ref{bc2}), and (\ref{newomg}). We use $\beta=-1.1$ for all plots, and take $\delta=0.01,\,\omega=0.8$ and $M/\sqrt{\phi}=1$ for Eq.(\ref{bc2}) and $\delta=0.01,\,\omega=0.8$ and ${\cal Q}=1$ for Eq.(\ref{newomg}).

\section{Energy Conditions}\label{En}
We can further explore and check the energy conditions. In this section, we only consider the two types of energy conditions to examine the wormhole solutions. The first one is null energy condition (NEC) given as $T_{\mu\nu}k^{\mu}k^{\nu}\geq 0$, which determines the non-negative value of energy-momentum tensor with $k_{\mu}$ being null vectors. The NEC yields $\rho + p_r \geq 0$. Please note that the null energy condition (NEC) can be understood as the requirement for the energy of particles moving along a null geodesic, such as photons and massless particles, to remain non-negative. Additionally, the strong energy condition (SEC) defined as $\left( T_{\mu\nu} - \frac{1}{2} T g_{\mu\nu} \right) X^{\mu} X^{\nu} \ge 0$ with $X_{\mu}$ being a timelike vector field, yields $
\rho + 2p_{t} \geq 0 $ and $\rho + p_{r} + 2p_{t} \geq 0 $. However, the traversable wormholes in some particular models, e.g., Casimir wormholes \cite{Garattini:2019ivd,Jusufi:2020rpw}, need the (exotic) matter which violates the energy conditions.

\subsection{Constants $\omega$ \& $\beta$}

\begin{figure}[ht!]
    \centering
    \includegraphics[width=0.48\textwidth]{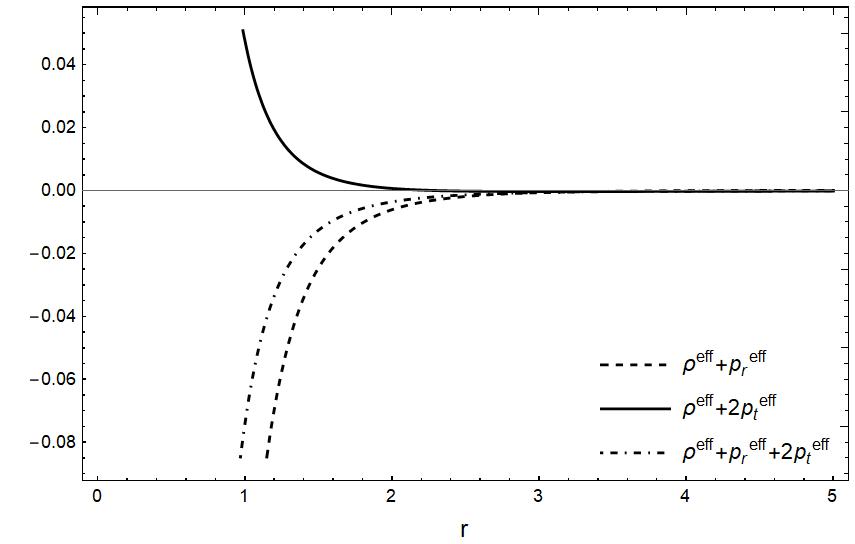}
    \caption{We show the behaviors of $\rho^{\text{eff}} +  p_r^{\text{eff}} $, $\rho^{\text{eff}} + 2 p_t^{\text{eff}} $ and $\rho^{\text{eff}}  + p_r^{\text{eff}} +2p_t^{\text{eff}}  $ versus $r$ using $\beta=-1.1$,  $r_0 =1$  $\omega =0.8 $ and ${\cal Q}=1$.}
    \label{weakno}
\end{figure}

We first consider constants $\omega$ and $\beta$ and take $b(r)$ given in Eq.(\ref{bc1}), and then compute the energy density. We find
\beq
\theta\,\rho(Y) &=& -\frac{1}{8 \pi  (\omega +1) Y^2}\left(\frac{Y}{Y_{0}}\right)^{\frac{2 (\omega +1)}{\omega  (2 \beta -1)+1}}+\delta\frac{P}{(4\pi)^{3/2}}\exp\Big(-\frac{Y^{2}}{4}\Big)\,.\label{newrr}
\eeq
Using $\rho(Y)$, we can compute $p_{Y}$ and $p_{Y}$ to obtain
\beq
\theta\,p_{r}(Y) &=& -\frac{\omega}{8 \pi  (\omega +1) Y^2}\left(\frac{Y}{Y_{0}}\right)^{\frac{2 (\omega +1)}{\omega  (2 \beta -1)+1}}\nonumber\\&-&\delta\frac{P}{8 \pi ^{3/2}Y^3}\Big(4 \sqrt{\pi } \left(\text{erf}\left(\frac{Y_{0}}{2}\right)-\text{erf}\left(\frac{Y}{2}\right)\right)+\left(Y^2+4\right) Y e^{-\frac{Y^2}{4 }}-4 Y_{0} e^{-\frac{Y_{0}^2}{4 }}\Big)\,,\label{inpr}\\ 
\theta\,p_t(Y) &=& -\frac{\omega \beta}{8 \pi  (\omega +1) Y^2}\left(\frac{Y}{Y_{0}}\right)^{\frac{2 (\omega +1)}{\omega  (2 \beta -1)+1}}\nonumber\\&-&\delta \frac{P}{4 \pi  \theta  Y^3}\Big(\text{erf}\left(\frac{Y}{2}\right)-\text{erf}\left(\frac{Y_{0}}{2}\right)-\frac{e^{-\frac{Y^2}{4}} Y}{\sqrt{\pi }}+\frac{e^{-\frac{Y_{0}^2}{4}} \text{Y0}}{\sqrt{\pi }}\Big)\,,\label{inpr}
\eeq

where $Y\equiv r/\sqrt{\theta},\,P\equiv M/\sqrt{\theta}$. The stress-energy tensor (SET) can be computed using Eq.(\ref{3}). We next consider constants $\omega$ and $\beta$ and take $b(r)$ given in Eq.(\ref{bc2}), and then compute the energy density to obtain
\beq
\phi\,\rho(Z) &=& -\frac{1}{8 \pi  (\omega +1) Z^2}\left(\frac{Z}{Z_{0}}\right)^{\frac{2 (\omega +1)}{\omega  (2 \beta -1)+1}}+\delta \frac{Q}{\pi ^2 }\left(\frac{1}{\left(1+Z^2\right)^2}\right) \,.\label{newrr2}
\eeq
Using $\rho(Z)$, we can compute $p_{r}$ and $p_{t}$ to obtain
\beq
\phi\,p_{r}(Z) &=& -\frac{\omega}{8 \pi  (\omega +1) Z^2}\left(\frac{Z}{Z_{0}}\right)^{\frac{2 (\omega +1)}{\omega  (2 \beta -1)+1}}\nonumber\\&-&\delta \frac{Q}{\pi ^2 Z^3}\left(-\frac{Z+3 Z^3}{\left(1+Z^2\right)^2}+\frac{Z_0}{1+Z_0 ^2}+\text{tan}^{-1}(Z)-\text{tan}^{-1}(Z_0)\right) \,,\label{inpr2}\\ \phi\,p_t(Z) &=& -\frac{\omega \beta}{8 \pi  (\omega +1) Z^2}\left(\frac{Z}{Z_{0}}\right)^{\frac{2 (\omega +1)}{\omega  (2 \beta -1)+1}}\nonumber\\&-&\delta \frac{Q }{2 \pi ^2 Z^3}\left(\frac{Z}{1+Z^2}-\frac{Z_0}{1+Z_0^2}-\text{tan}^{-1}(Z)+\text{tan}^{-1}(Z_0)\right)  \,,\label{inpt2}
\eeq
where $Z\equiv r/\sqrt{\phi},\,Q\equiv M/\sqrt{\phi}$. We find from Fig.(\ref{fig:enter-label0em}) that for small values of $\delta$ the NEC and SEC cannot be satisfied. However, since $b(r)$ is dependent on $\delta$, the usual properties of $b(r)$ cannot be satisfied, e.g., $b(r)/r \nrightarrow 0$ for $r\rightarrow \infty$ if $\delta$ is not so small, $\delta\gg {\cal O}(1)$.  
\begin{figure}[ht!]
    \centering
\includegraphics[width=3in,height=3in,keepaspectratio=true]{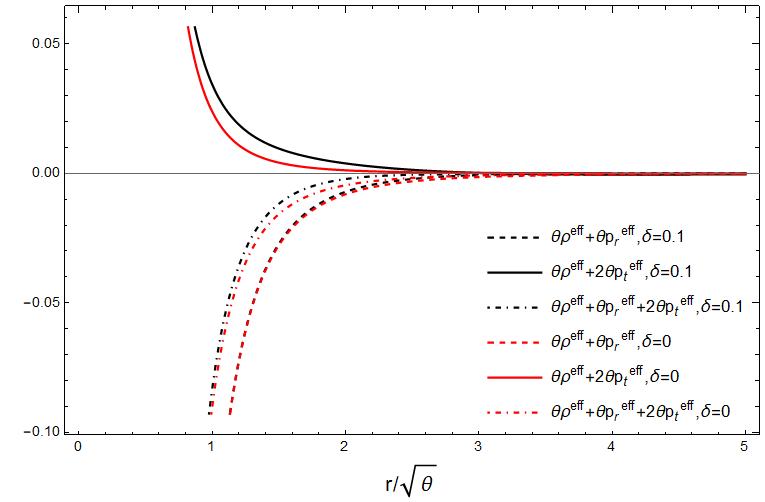}
\includegraphics[width=3in,height=3in,keepaspectratio=true]{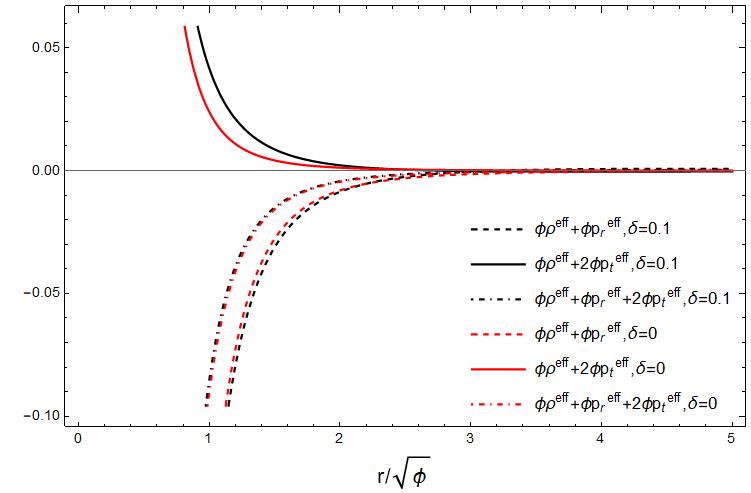}
    \caption{We show the behaviors of $\theta\rho^{\text{eff}} + \theta p_r^{\text{eff}}$, $\theta\rho^{\text{eff}} + 2\theta p_t^{\text{eff}}$, $\theta\rho^{\text{eff}} +\theta p_r^{\text{eff}}+2\theta p_t^{\text{eff}} $ versus $r/\sqrt{\theta}$ using $\beta=-1.1$,  $r_0 =1$, $\omega =0.8 $ for the Gaussian distribution Eq.(\ref{newb30}) (Left panel), and $\phi\rho^{\text{eff}} + \phi p_r^{\text{eff}}$, $\phi\rho^{\text{eff}} + 2\phi p_t^{\text{eff}}$, $\phi\rho^{\text{eff}} +\phi p_r^{\text{eff}}+2\phi p_t^{\text{eff}} $ versus $r/\sqrt{\phi}$ using $\beta=-1.1$,  $r_0 =1$, $\omega =0.8 $ for the Lorentzian distribution Eq.(\ref{newb3011}) (Right panel).}
    \label{fig:enter-label0no}
\end{figure}
We next consider constants $\omega$ and $\beta$ and take $b(r)$ given in Eq.(\ref{bem}), and then compute the energy density. We find
\beq
\rho(r) &=& -\frac{1}{8 \pi  (\omega +1) r^2}\left(\frac{r}{r_{0}}\right)^{\frac{2 (\omega +1)}{\omega  (2 \beta -1)+1}}+\delta\frac{{\cal Q}^{2}}{8\pi r^{4}}\,.\label{newrrem}
\eeq
Using $f(r)$, we can compute $p_{r}=\theta^{1}_{1}$ and $p_{t}=\theta^{2}_{2}$ to obtain
\beq
p_{r}(r) &=& -\frac{\omega}{8 \pi  (\omega +1) r^2}\left(\frac{r}{r_{0}}\right)^{\frac{2 (\omega +1)}{\omega  (2 \beta -1)+1}}-\delta\frac{{\cal Q}^2}{8 \pi  r^4}\Big(\frac{2 r}{r_{0}}-3\Big)\,,\label{inpr}\\ 
p_t(r) &=& -\frac{\omega\beta}{8 \pi  (\omega +1) r^2}\left(\frac{r}{r_{0}}\right)^{\frac{2 (\omega +1)}{\omega  (2 \beta -1)+1}}-\delta \frac{{\cal Q}^2}{8 \pi  r^4}\Big(1-\frac{r}{r_{0}}\Big)\,.\label{inpr}
\eeq
The behaviors of $\theta\rho^{\text{eff}} + \theta p_r^{\text{eff}}$, $\theta\rho^{\text{eff}} + 2\theta p_t^{\text{eff}}$, $\theta\rho^{\text{eff}} +\theta p_r^{\text{eff}}+2\theta p_t^{\text{eff}} $ versus $r/\sqrt{\theta}$ for the Gaussian distribution Eq.(\ref{newb30}), and $\phi\rho^{\text{eff}} + \phi p_r^{\text{eff}}$, $\phi\rho^{\text{eff}} + 2\phi p_t^{\text{eff}}$, $\phi\rho^{\text{eff}} +\phi p_r^{\text{eff}}+2\phi p_t^{\text{eff}} $ versus $r/\sqrt{\phi}$ for the Lorentzian distribution Eq.(\ref{newb3011}) has been illustrated in Fig.(\ref{fig:enter-label0no}). In Fig.(\ref{weakno}), we notice that the NEC is also violated for the charged fluid deformation.

\subsection{Variable $\omega$ \& Constant $\beta$}
In this case, we can easily compute the energy density to obtain
\beq
\theta\,\rho(Y) &=& -\frac{\beta }{4 \pi  (\beta -1) Y^2}\left(\frac{Y}{Y_{0}}\right)^{\frac{2}{2 \beta -1}+1} \left(\frac{\beta  Y-2 \beta  Y_{0}+r_{0}}{Y_{0}-\beta Y_{0}}\right)^{\frac{2}{1-2 \beta }+1}+\delta\frac{P}{(4\pi)^{3/2}}\exp\Big(-\frac{Y^{2}}{4}\Big)\,.\label{newrro}
\eeq
Using $\rho(Y)$, we can compute $p_{Y}$ and $p_{Y}$ to obtain
\beq
\theta\,p_{r}(Y) &=& \frac{1}{4 \pi  (\beta -1) Y^2}\left(\frac{Y}{Y_{0}}\right)^{\frac{2}{2 \beta -1}} \left(\frac{\beta  Y-2 \beta Y_{0}+Y_{0}}{Y_{0}-\beta Y_{0}}\right)^{\frac{2}{1-2 \beta }+1}\nonumber\\&-&\delta\frac{P}{8 \pi ^{3/2}Y^3}\Big(4 \sqrt{\pi } \left(\text{erf}\left(\frac{Y_{0}}{2}\right)-\text{erf}\left(\frac{Y}{2}\right)\right)+\left(Y^2+4\right) Y e^{-\frac{Y^2}{4 }}-4 Y_{0} e^{-\frac{Y_{0}^2}{4 }}\Big)\,,\label{inpro}\\ 
\theta\,p_t(Y) &=& \frac{\beta}{4 \pi  (\beta -1) Y^2}\left(\frac{Y}{Y_{0}}\right)^{\frac{2}{2 \beta -1}} \left(\frac{\beta  Y-2 \beta Y_{0}+Y_{0}}{Y_{0}-\beta Y_{0}}\right)^{\frac{2}{1-2 \beta }+1}\nonumber\\&-&\delta \frac{P}{4 \pi  \theta  Y^3}\Big(\text{erf}\left(\frac{Y}{2}\right)-\text{erf}\left(\frac{Y_{0}}{2}\right)-\frac{e^{-\frac{Y^2}{4}} Y}{\sqrt{\pi }}+\frac{e^{-\frac{Y_{0}^2}{4}} \text{Y0}}{\sqrt{\pi }}\Big)\,,\label{inpro}
\eeq
We next consider $\omega(r)$ and constant $\beta$ and take $b(r)$ given in Eq.(\ref{newomg}), and then compute the energy density to obtain
\beq
\phi\,\rho(Z) &=& -\frac{\beta }{4 \pi  (\beta -1) Z^2}\left(\frac{Z}{Z_{0}}\right)^{\frac{2}{2 \beta -1}+1} \left(\frac{\beta  Z-2 \beta  Z_{0}+Z_{0}}{Z_{0}-\beta Z_{0}}\right)^{\frac{2}{1-2 \beta }+1}+\delta \frac{Q}{\pi ^2 }\left(\frac{1}{\left(1+Z^2\right)^2}\right) \,.\label{newrr2o}
\eeq
Using $\rho(Z)$, we can compute $p_{r}$ and $p_{t}$ to obtain
\beq
\phi\,p_{r}(Z) &=& \frac{1}{4 \pi  (\beta -1) Z^2}\left(\frac{Z}{Z_{0}}\right)^{\frac{2}{2 \beta -1}} \left(\frac{\beta  Z-2 \beta Z_{0}+Z_{0}}{Z_{0}-\beta Z_{0}}\right)^{\frac{2}{1-2 \beta }+1}\nonumber\\&-&\delta \frac{Q}{\pi ^2 Z^3}\left(-\frac{Z+3 Z^3}{\left(1+Z^2\right)^2}+\frac{Z_0}{1+Z_0 ^2}+\text{tan}^{-1}(Z)-\text{tan}^{-1}(Z_0)\right) \,,\label{inpr2o}\\ \phi\,p_t(Z) &=& \frac{\beta}{4 \pi  (\beta -1) Z^2}\left(\frac{Z}{Z_{0}}\right)^{\frac{2}{2 \beta -1}} \left(\frac{\beta  Z-2 \beta Z_{0}+Z_{0}}{Z_{0}-\beta Z_{0}}\right)^{\frac{2}{1-2 \beta }+1}\nonumber\\&-&\delta \frac{Q}{2 \pi ^2 Z^3}\left(\frac{Z}{1+Z^2}-\frac{Z_0}{1+Z_0^2}-\text{tan}^{-1}(Z)+\text{tan}^{-1}(Z_0)\right)  \,.\label{inpt2o}
\eeq
\begin{figure}[ht!]
    \centering
\includegraphics[width=3in,height=3in,keepaspectratio=true]{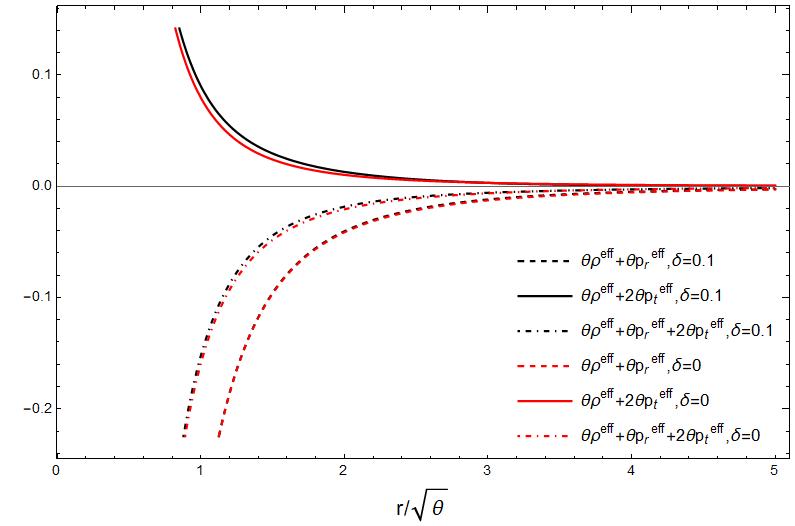}
\includegraphics[width=3in,height=3in,keepaspectratio=true]{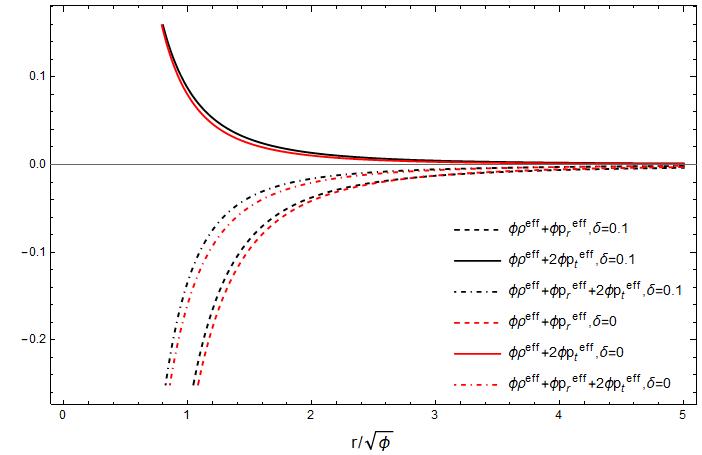}
    \caption{We show the behaviors of $\theta\rho^{\text{eff}} + \theta p_r^{\text{eff}}$, $\theta\rho^{\text{eff}} + 2\theta p_t^{\text{eff}}$, $\theta\rho^{\text{eff}} +\theta p_r^{\text{eff}}+2\theta p_t^{\text{eff}} $ versus $r/\sqrt{\theta}$ using $\beta=-0.01$,  $r_0 =1$, $P=1$ for the Gaussian distribution Eq.(\ref{newb30}) (Left panel), and $\phi\rho^{\text{eff}} + \phi p_r^{\text{eff}}$, $\phi\rho^{\text{eff}} + 2\phi p_t^{\text{eff}}$, $\phi\rho^{\text{eff}} +\phi p_r^{\text{eff}}+2\phi p_t^{\text{eff}} $ versus $r/\sqrt{\phi}$ using $\beta=-0.01$,  $r_0 =1$, $ Q =1$ for the Lorentzian distribution Eq.(\ref{newb3011}) (Right panel).}
    \label{fig:enter-label0n}
\end{figure}
We consider $\omega(r)$ and constant $\beta$ and take $b(r)$ given in Eq.(\ref{newomg}), and then compute the energy density. We find
\beq
\rho(r) &=& -\frac{\beta }{4 \pi  (\beta -1) r^2}\left(\frac{r}{r_{0}}\right)^{\frac{2}{2 \beta -1}+1} \left(\frac{\beta  r-2 \beta  r_{0}+r_{0}}{r_{0}-\beta r_{0}}\right)^{\frac{2}{1-2 \beta }+1}+\delta\frac{{\cal Q}^{2}}{8\pi r^{4}}\,.\label{newrremoo}
\eeq
Using $f(r)$, we can compute $p_{r}=\theta^{1}_{1}$ and $p_{t}=\theta^{2}_{2}$ to obtain
\beq
p_{r}(r) &=& \frac{1}{4 \pi  (\beta -1) r^2}\left(\frac{r}{r_{0}}\right)^{\frac{2}{2 \beta -1}} \left(\frac{\beta  r-2 \beta r_{0}+r_{0}}{r_{0}-\beta r_{0}}\right)^{\frac{2}{1-2 \beta }+1}-\delta\frac{{\cal Q}^2}{8 \pi  r^4}\Big(\frac{2 r}{r_{0}}-3\Big)\,,\label{inproo}\\ 
p_t(r) &=& \frac{\beta}{4 \pi  (\beta -1) r^2}\left(\frac{r}{r_{0}}\right)^{\frac{2}{2 \beta -1}} \left(\frac{\beta  r-2 \beta r_{0}+r_{0}}{r_{0}-\beta r_{0}}\right)^{\frac{2}{1-2 \beta }+1}-\delta \frac{{\cal Q}^2}{8 \pi  r^4}\Big(1-\frac{r}{r_{0}}\Big)\,,\label{inproo}
\eeq
\begin{figure}[ht!]
    \centering
    \includegraphics[width=0.48\textwidth]{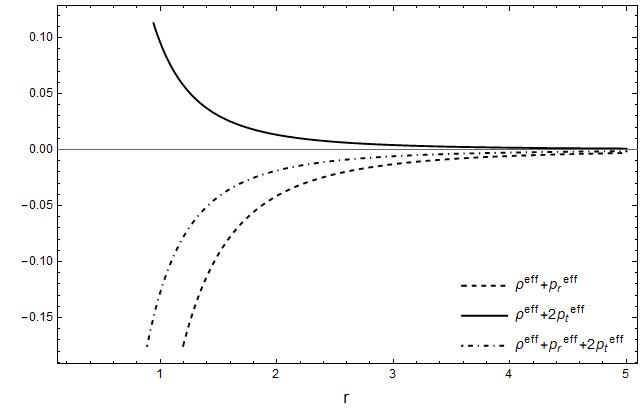}
    \caption{We show the behaviors of $\rho^{\text{eff}} +  p_r^{\text{eff}} $, $\rho^{\text{eff}} + 2 p_t^{\text{eff}} $ and $\rho^{\text{eff}}  + p_r^{\text{eff}} +2p_t^{\text{eff}}  $ versus $r$ using $\beta=-0.01$,  $r_0 =1$  $\delta =0.01 $and ${\cal Q}=1$. }
    \label{weaknn}
\end{figure}
The behaviors of $\theta\rho^{\text{eff}} + \theta p_r^{\text{eff}}$, $\theta\rho^{\text{eff}} + 2\theta p_t^{\text{eff}}$, $\theta\rho^{\text{eff}} +\theta p_r^{\text{eff}}+2\theta p_t^{\text{eff}} $ versus $r/\sqrt{\theta}$ for the Gaussian distribution Eq.(\ref{newb30}) and $\phi\rho^{\text{eff}} + \phi p_r^{\text{eff}}$, $\phi\rho^{\text{eff}} + 2\phi p_t^{\text{eff}}$, $\phi\rho^{\text{eff}} +\phi p_r^{\text{eff}}+2\phi p_t^{\text{eff}} $ versus $r/\sqrt{\phi}$ for the Lorentzian distribution Eq.(\ref{newb3011}) can be seen in Fig.(\ref{fig:enter-label0n}). In Fig.(\ref{weaknn}), we notice that the NEC is also violated for the charged fluid deformation.

\section{Weak gravitational lensing}\label{lens}

Embarking on this section, we revisit the Gauss-Bonnet theorem and embark on the calculation of the weak deflection angle for wormhole configurations. Our starting point is the expression for null geodesics, $ds^2=0$, which can be rearranged to yield:
\begin{eqnarray}
dt^2=\gamma_{ij}dx^i dx^j=\frac{1}{W}dr^2+r^2 d\Omega^2,~\label{opmetric}
\end{eqnarray}

Within this context, indices i and j represent the spatial dimensions  $(1 to 3) $, and $\gamma_{ij}$ denotes the optical metric. To utilize the Gauss-Bonnet theorem effectively, we must first calculate the Gaussian curvature associated with Eq.(\ref{eqn46}). This calculation is presented in detail here:
\begin{eqnarray}
\mathcal{K}=\frac{R}{2}&\approx& \frac{6 \beta ^2 r_0^2}{r^4}+\frac{2 \beta  r_0^2}{r^4}-\frac{r_0^2}{r^4}-\frac{6 \beta ^2 r_0}{r^3}-\frac{2 \beta  r_0}{r^3}.~\label{GC}
\end{eqnarray}
Within this framework, $\gamma\equiv\det (\gamma_{ij})$ represents the determinant of the optical metric, and $R$ denotes the Ricci scalar. Let $D$ be a compact, oriented, nonsingular two-dimensional Riemannian surface characterized by its Euler characteristic $\chi(D)$and Gaussian curvature $\mathcal{K}$. This domain is enclosed by a piecewise-smooth curve with geodesic curvature $\kappa$. The link between the deflection angle of light and the Gaussian curvature stems from the Gauss-Bonnet theorem, which is invoked by employing the following expressions:
\begin{eqnarray}
\int\int_D \mathcal{K}dS+\oint_{\partial D}\kappa dt+\sum_{i=1}\beta_i=2\pi \chi(D),~\label{GB}
\end{eqnarray}

Within this context, $dS$ represents the differential element of area, $\kappa$ denotes the geodesic curvature of the boundary, defined as $\kappa=|\nabla_{\dot{C}}\dot{C}|$ and $\beta_i$ represents the $i^{\text{th}}$ exterior angle. For a particular region  
$\tilde{D}$ enclosed by a geodesic $C_1$ extending from the source $S$ to the observer $O$ and a circular curve $C_R$ intersecting $C_1$ orthogonally at $S$ and $O$, Equation (\ref{GB}) reduces to:
\begin{eqnarray}
\int\int_{\tilde{D}}\mathcal{K}dS+\int_{C_R}\kappa(C_R)dt=\pi,~\label{GB2}
\end{eqnarray}
During this derivation, we employed the conditions $\kappa(C_1)=0$ and the Euler characteristic $\chi(\tilde{D})=1$. For the specific circular curve $C_R := r(\phi)=R=\text{const}$, the non-zero segment of the geodesic curvature is calculated as:
\begin{eqnarray}
\kappa(C_R)=\left(\nabla_{\dot{C}_R}\dot{C}_R\right)^r=\dot{C}^{\phi}_R(\partial_{\phi}\dot{C}^r_R)+\Gamma^r_{\phi\phi}(\dot{C}^{\phi}_R)^2,
\end{eqnarray}

In this context, $\dot{C}_R$ represents the directional derivative of the circular curve  $C_R$, and $\Gamma^r_{\phi\phi}$ signifies the Christoffel symbol corresponding to the optical metric (\ref{opmetric}). As $R$ tends towards infinity, we arrive at:

\begin{eqnarray}
&&\lim_{R\rightarrow \infty}\left[\kappa(C_R)dt\right]=d\phi.~\label{geoR}
\end{eqnarray}

\begin{figure}[ht!]
    \centering
\includegraphics[width=0.6\textwidth]{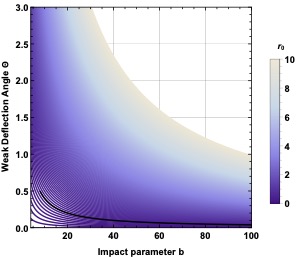}
    \caption{The deflection angle, $\Theta$, as expressed in Eq.(\ref{deflang}) is plotted against the impact parameter $b$ using $\beta=-1.1$ in Eq.(\ref{eqn46}). }
    \label{weakoo}
\end{figure}

By substituting Eq.~(\ref{geoR}) into Eq.~(\ref{GB2}), we arrive at:
\begin{eqnarray}
\int\int_{D}\mathcal{K}dS+\int_0^{\pi+\Theta} d\phi\pi=\pi.
\end{eqnarray}
In this context, the surface area on the equatorial plane is formulated as \cite{Gibbons:2008rj}:
\begin{equation}
dS=\sqrt\gamma dr d \phi
\label{}
\end{equation}
Following this, the weak deflection angle of light can be determined for Eq.\ref{eqn46} as:
\begin{eqnarray}
\Theta&=&-\int\int_{\tilde{D}}\mathcal{K}dS=-\int^{\pi}_0\int^{\infty}_{\frac{b}{\sin\phi}}\mathcal{K}dS \nonumber\\
&\simeq&-\frac{3 \pi  \beta ^2 r_0^2}{2 b^2}-\frac{\pi  \beta  r_0^2}{2 b^2}+\frac{\pi  r_0^2}{4 b^2}+\frac{12 \beta ^2 r_0}{b}+\frac{4 \beta  r_0}{b}.~\label{deflang}
\end{eqnarray}

In this analysis, we utilized the zero-order particle trajectory $r=b/\sin\phi$, where $0\leq\phi\leq\pi$in the weak deflection limit. The dependence of the deflection angle on the impact parameter, as influenced by the wormhole geometry, is presented graphically in Fig.(\ref{weakoo}). Our findings reveal that the deflection angle is contingent upon the parameters $\beta$ and $r_{0}$ for $\omega(r)=-\tfrac{r_{0}}{\beta r}$. For specific values of $\beta$, it is observed that the deflection angle increases as $r_{0}$ grows. These results warrant comparison with those obtained using an alternative approach recently proposed in Ref.\cite{Li:2024ugi}.

\section{Concluding Remarks}

In this work, we investigate wormhole solutions invoking the Minimal Geometric Deformation (MGD) procedure within the framework of Unimodular Gravity. We employed a static and spherically symmetric Morris-Thorne traversable wormholes and considered both constants and variables in the equation of state parameter. More specifically, we considered $\omega={\rm cont.}$ and $\omega=\omega(r)=-\tfrac{r_{0}}{\beta r}$.  We computed field equations and derived the shape functions. We showed that the usual properties of the obtained shape functions has been satisfied. We considered the gravitational decoupling technique and considered various forms of the energy density for a smeared and particle-like gravitational source in the context of noncommutative geometry and a statically charged fluid. We obtained the novel wormhole solutions showing the violation of the Null Energy Condition (NEC). We used the Gauss-Bonnet theorem to compute the weak deflection angle of light for the wormhole solutions. We also found that the deflection angle depends upon the parameter $\beta$ and $r_{0}$ for $\omega(r)=-\tfrac{r_{0}}{\beta r}$. 

We can further test whether these wormholes be sustained by their own quantum fluctuations. In particular, the energy density of the graviton-one loop contribution to classical energy in a traversable wormhole background and the finite one loop energy density have to be considered as a self-consistent source for these wormhole geometries. To this end, we shall follow the existing publications, see e.g., \cite{Garattini:2005gd,Garattini:2007ff,Garattini:2008xz}. However, the wormhole equation of state is still unknown and hence the present work can be tested by the wormhole observations, see. e.g., \cite{Abe:2010ap,Toki:2011zu,Godani:2021aub,Ohgami:2015nra,Dai:2019mse}.

\acknowledgments
P. P. is financially supported by a DPST scholarship for her undergraduate study. N.K. is funded by the National Research Council of Thailand (NRCT): Contract number N42A660971. P.C. is financially supported by Thailand NSRF via PMU-B under grant number PCB37G6600138. A. {\"O}. would like to acknowledge the contribution of the COST Action CA21106 - COSMIC WISPers in the Dark Universe: Theory, astrophysics and experiments (CosmicWISPers) and the COST Action CA22113 - Fundamental challenges in theoretical physics (THEORY-CHALLENGES). We also thank TUBITAK and SCOAP3 for their support.

\end{document}